\documentclass[aps,prl,floatfix,groupedaddress,,nofootinbib,notitlepage,twocolumn]{revtex4-2}
\usepackage{times,bm,bbm,bbold,amssymb,amsmath,amsfonts,dsfont,cancel,graphics,graphicx,color}
\usepackage[normalem]{ulem}
\usepackage[usenames,dvipsnames]{xcolor}
\usepackage[colorlinks=true,citecolor=blue,linkcolor=blue,urlcolor=blue]{hyperref}

\DeclareFontFamily{OT1}{pzc}{}
\DeclareFontShape{OT1}{pzc}{m}{it}
              {<-> s * [1.25] pzcmi7t}{}
\DeclareMathAlphabet{\mathpzc}{OT1}{pzc}
                                 {m}{it}
\usepackage[T1]{fontenc} 
\usepackage{newtxtext,newtxmath} 

\def \[[{{[\hskip-1mm[}}
\def \]]{{]\hskip-0.9mm]}}

\begin{document}
%\pagestyle{fancy}
%\chead[DRAFT]{DRAFT: For exclusive use in IP protection}
%\lhead[]{}
%\rhead[]{}
\title{State-Based Quantum Operations: Chameleon Gates}

\author{S. Alipour}
\email{s.alipoor@gmail.com}
\affiliation{Chameleon Computing, 02630 Espoo, Finland}

\author{A. T. Rezakhani}
%\email{ali_rezakhani@yahoo.com}
\affiliation{Chameleon Computing, 02630 Espoo, Finland}

%%%%%%%%%%%%%%%%%%%%%%%%%%%%%%%%%%%%%%%%%%%%%%%%%%%%%%%%%%%%%%%%
\begin{abstract}
We introduce \textit{chameleon gates} as a natural generalization of conventional quantum controlled-gates. Chameleon gates are \textit{state-based} quantum controlled-operations that retain standard elements such as control and target systems, while introducing a new feature: the \textit{quantum knob}. This knob is a quantum signal (state) that determines the operation performed by the gate. Consequently, the action and form of a chameleon gate depend dynamically on the quantum knob, allowing the gate to adapt its operation and implement transformations that are not necessarily unitary. This \textit{shapeshifting} property is in stark contrast to conventional quantum controlled-gates, whose actions are fixed and cannot be modified. We also propose how chameleon gates can be realized using conventional quantum gates available in current quantum technologies. We then employ chameleon gates as a useful building block within the recently proposed state-based quantum computation (SBQC) framework. Using this approach, we demonstrate the simulation of state-dependent (nonlinear) quantum evolutions. 
\end{abstract}
\date{\today}
\maketitle
%\thispagestyle{fancy}
%%%%%%%%%%%%%%%%%%%%%%%%%%%%%%%%%%%%%%%%%%%%%%%%%%%%%%%%%%%%%%%%

%%%%%%%%%%%%%%%%%%%%%%%%%%%%%%%%%%%%%%%%%%%%%%%%%%%%%%%%%%%%%%%%
%\section{Introduction}

\textit{Introduction.---}In the standard paradigm of quantum computation, gates play a pivotal role as the building blocks of quantum algorithms \cite{book:Nielsen-Chuang}. Quantum gates are unitary operators which have fixed actions on any input state. Examples of such operations include single-qubit gates such as Hadamard gate, $X$-, $Y$-, and $Z$-Pauli gates, phase gate $e^{-i\pi Z/4}$, and two-qubit gates such as \textsc{swap} and perfect entangling gates \cite{Ali-SPE}. An important class of quantum gates are controlled-$U$ gates, which apply a fixed unitary operator on a target system or do nothing conditioned on the state of another control system. This type of gates includes quantum gates such as controlled-\textsc{not} (\textsc{cnot}), controlled-phase (\textsc{cz}), and controlled-\textsc{swap} (Fredkin) operations, which are of paramount importance in universal quantum computation techniques including circuit-based, measurement-based, and state-based quantum computation \cite{SBQS}. Methods for devising controlled-$U$ gate from any given gate $U$ have already been suggested in the literate \cite{O'Brien:adding-control}. 

Despite their enormous utility, that in controlled gates the applied operation is fixed and predetermind can make simulation of some complex physical and computational processes difficult. Here we extend fixed gates to a new type of quantum gates with varying action which is prescribed by an input state (``quantum knob'') to the gate.  Due to this varying behavior, which changes according to incoming quantum states or signal, we refer to these gates as the ``chameleon gates.'' These state-dependent gates are used as fundamental building blocks of the state-based quantum computation paradigm.  

%\section{State-based quantum operations}

\begin{figure}[bp]
\includegraphics[width=\linewidth]{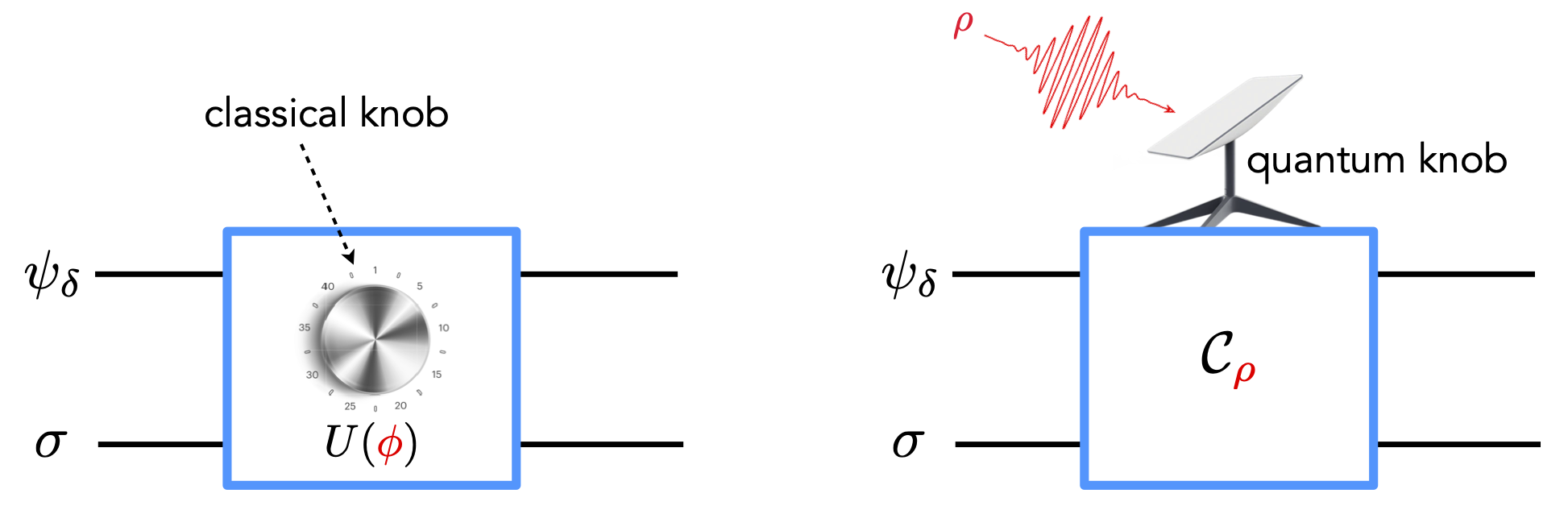}
\caption{Comparing conventional parameter-dependent quantum gates controlled-$U(\phi)$ with operator-dependent quantum controlled-gates $\mathcal{C}_{\rho}$. Both gates are controlled by some \textit{knobs}; while the knob in the left one is \textit{classical} (a parameter $\phi$), the knob in the right one is \textit{quantum} (an operator, a quantum state/signal $\rho$). Both controlled gates have standard elements such as control and target systems, which here are represented by $\psi_{\delta}$ and $\sigma$. For the specific definitions of $\psi_{\delta}$ and $\mathcal{C}_{\rho}$ see the main text.}
\label{fig:knobs}
\end{figure}

\textit{State-based quantum operations.---}One way to enhance quantum controlled-gates to be more versatile is to make them parameter dependent, i.e., defining controlled-$U(\phi)$ gates $|0\rangle\langle 0|\otimes \openone+|1 \rangle \langle 1|\otimes U(\phi)$, where $\phi$ is a tunable parameter of the quantum circuit. An example of such gates is the controlled-phase gate, where $\phi$ is a phase shift and $U(\phi)=e^{i\phi Z}$. The parameter $\phi$ acts as a classical signal or a \textit{classical knob} that modifies the values of the elements of the gate---Fig. \ref{fig:knobs} [left].

%%%%%%%%%%%%%%%%%%%%%%%%%%%%%%%%%%%%%%%%%%%%%%%%%%%%%%%%%%%%%%%%
%%%%%%%%%%%%%%%%%%%%%%%%%%%%%%%%%%%%%%%%%%%%%%%%%%%%%%%%%%%%%%%%

A yet further, natural generalization of such parametrized controlled-gates can be achieved by replacing the scalar parameter $\phi$ by a Hermitian \textit{operator} $A$ such that the unitary quantum gate becomes $U(A)$. This kind of generalization leads, for example, to the extension of the geometric phase to non-Abelian holonomic phases, where $U(A)=e^{i A}$ \cite{nonAbelian-phase}. 

Here we introduce a novel class of quantum controlled-gates where the operator $A$ is replaced with a \textit{quantum state} $\rho$, which represents a quantum signal acting as a \textit{quantum knob} for the gate. That is, these gates are \textit{state-based} quantum operations whose form are specified by a quantum signal or knob state---Fig. \ref{fig:knobs} [right]. We later show that this replacement enables utilization of the power of quantum states along with quantum operations in developing new quantum algorithms. The key observation is that in addition to using fixed quantum states as the quantum knobs, we can also use any quantum state from the very quantum circuit itself (or part of the circuit) or we can even use the state of another quantum circuit. This possibility significantly expands the domain of computations from linear dynamics on quantum computers to nonlinear evolutions.  

In the following, we start with introducing \textit{state-based} controlled-gates and later %in Sec. \ref{sec:SBQC}
use them to do state-based quantum computation.\\

%%%%%%%%%%%%%%%%%%%%%%%%%%%%%%%%%%%%%%%%%%%%%%%%%%%%%%%%%%%%%%%%
\textit{State-based controlled-gates.}---The action of a state-based controlled-gate $\mathpzc{U}_{\rho}$, which is not necessarily a unitary gate, resembles an ordinary controlled-$\rho$ gate if $\rho$ were a unitary operator. This means that such a quantum gate has the ability to apply a function of the operator $\rho$, where $\rho$ is some quantum state rather than a unitary operation, on any given input target state $\sigma$ in a controlled fashion. Although we can choose any function here, in the following we make this simplification that $\rho$ itself acts on the target. If the control qubit is in the state $|0 \rangle$, the input target state $\sigma$ remains unchanged, whereas if the control qubit is $|1 \rangle$, the operation $\rho$ is applied on $\sigma$. This means that the action of $\mathpzc{U}_{\rho}$ can be described by a bipartite operation as 
\begin{align}
\mathpzc{U}_{\rho}= |0 \rangle\langle 0|\otimes \openone + |1 \rangle\langle 1|\otimes {\rho}.
\label{c-U-r}
\end{align}
This implies that if the control qubit is in the state 
\begin{align}
\psi_{\delta}=|\psi_{\delta}\rangle\langle \psi_{\delta}|,
\label{control-qubit}
\end{align}
where $|\psi_{\delta}\rangle= |0 \rangle + \delta |1 \rangle$, and we apply $\mathpzc{U}_{\rho}$ on this control qubit and the target system with state $\sigma$, we obtain
\begin{align}
\label{controlled-rho}
\mathpzc{U}_{\rho} (\psi_{\delta} \otimes \sigma)\mathpzc{U}^{\dag}_{\rho}=&\, |0 \rangle\langle 0| \otimes \sigma + \delta |1 \rangle\langle 0| \otimes \rho\sigma \\
&\,+ \delta^{*} |0 \rangle\langle 1| \otimes \sigma\rho + |\delta|^{2} |1 \rangle\langle 1| \otimes \rho \sigma \rho. \nonumber
\end{align}
The reason we use this particular $\psi_{\delta}$ as the state of the control qubit is for later convenience.
\\

%%%%%%%%%%%%%%%%%%%%%%%%%%%%%%%%%%%%%%%%%%%%%%%%%%%%%%%%%%%%%%%%
\begin{figure}[tp]
\includegraphics[width=\linewidth]{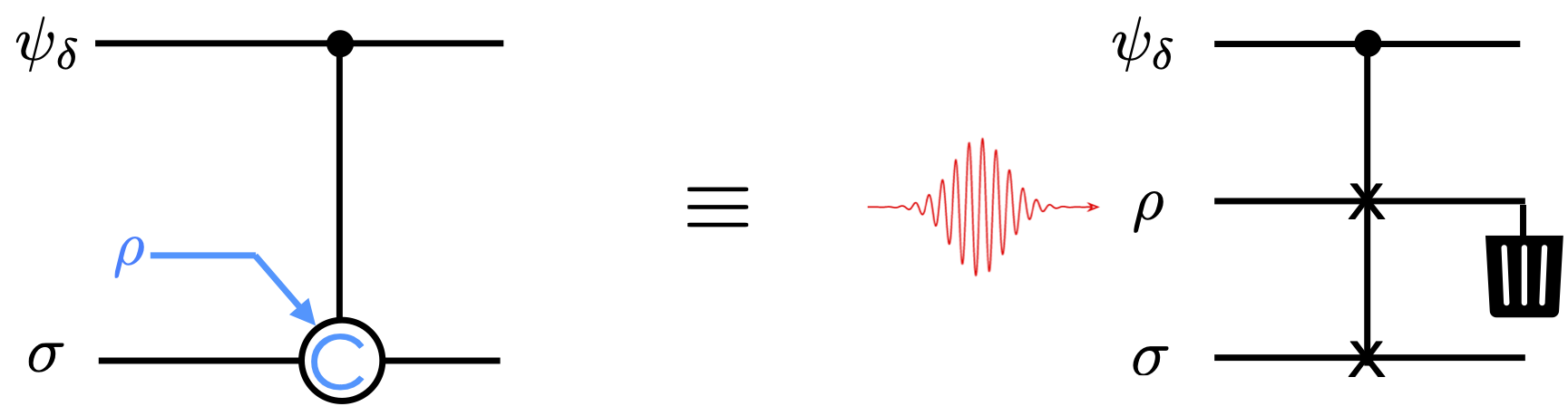}
\caption{[Left] Quantum-circuit schematic of a state-based controlled-gate $\mathcal{C}_{\rho}$, which we refer to as the chameleon gate [Eq. (\ref{chameleon-gate})]. [Right] An equivalent standard gate-based construction is also shown for this gate, which uses a specific combination of quantum states, quantum controlled-\textsc{swap} gate, and partial tracing. Note that basically the control qubit can be any state. But when the control qubit is chosen as $\psi_{\delta}$ with $|\delta|\ll 1$, this also yields an implementation for $\mathpzc{U}_{\rho}$ [Eq. (\ref{c-U-r})] to an $O(\delta^{2})$ error.}
\label{fig:equivalence}
\end{figure}
%%%%%%%%%%%%%%%%%%%%%%%%%%%%%%%%%%%%%%%%%%%%%%%%%%%%%%%%%%%%%%%%

%%%%%%%%%%%%%%%%%%%%%%%%%%%%%%%%%%%%%%%%%%%%%%%%%%%%%%%%%%%%%%%%
\textit{Realization of state-based controlled-gates.---}Here we show how we can realize a state-based operation using basic elements of the standard gate-based quantum computation model. When $\delta \ll 1$ the effect of  $\mathpzc{U}_{\rho}$ on the control qubit and the target system $\sigma$ can be realized as follows:

\begin{itemize}

\item Prepare an ancilla qubit (which will be used later as a control qubit) in the state $\psi_{\delta}$ [Eq. \eqref{control-qubit}] with $\delta \ll 1$.

\item Prepare a quantum system in the state $\rho$ and consider the input target system in the state $\sigma$.

\item Use the ancilla as the control system and the system with state $\rho$ as the first target system $\textsc{t}_1$ and the system with state $\sigma$ as the second target system $\textsc{t}_2$ and apply a controlled-\textsc{swap} gate $U_{\mathrm{cs}}$ on the control system and the two target systems:
\begin{align}
&U_{\mathrm{cs}} \left( \psi_{\delta} \otimes \rho \otimes  \sigma\right) U_{\mathrm{cs}}^{\dag} = |0 \rangle\langle 0| \otimes \rho \otimes \sigma  + \delta |1 \rangle\langle 0| \otimes  S (\rho \otimes \sigma)\nonumber\\
&~~~\,\,+ \delta^{*} |0 \rangle\langle 1| \otimes  (\sigma \otimes  \rho) S + |\delta|^{2} |1 \rangle\langle 1| \otimes S (\rho \otimes  \sigma) S,
\end{align}
where $S$ is the $\textsc{swap}$ operation, defined as $S(A\otimes B)S =B\otimes A$, $\forall A,B$.

\item Partial trace over the first target system $\textsc{t}_1$: 
\begin{align}
\label{eq.4}
 \mathrm{Tr}_{\textsc{t}_1} \big[ U_{\mathrm{cs}} \big(\psi_{\delta} \otimes \rho \otimes  &\, \sigma\big) U_{\mathrm{cs}}^{\dag} \big] = |0 \rangle\langle 0| \otimes  \sigma + \delta |1 \rangle\langle 0| \otimes  \rho \sigma \nonumber\\
&\,+ \delta^{*} |0 \rangle\langle 1| \otimes \sigma \rho + |\delta|^{2} |1 \rangle\langle 1| \otimes  \rho.
\end{align}
The final result of the set of instructions above is equivalent to the action of a state-based controlled-gate $\mathpzc{U}_{\rho}$ as in Eq. \eqref{controlled-rho} up to $O(\delta^{2})$, 
\begin{equation*}
\mathrm{Tr}_{\textsc{t}_1} \big[ U_{\mathrm{cs}} \left( \psi_{\delta} \otimes \rho \otimes  \sigma\right) U_{\mathrm{cs}}^{\dag} \big]= \mathpzc{U}_{\rho} (\psi_{\delta}\otimes \sigma)\mathpzc{U}_{\rho} ^{\dag}+ O(\delta^{2}).
\end{equation*}
Figure \ref{fig:equivalence} depicts these steps. In this figure we also introduce a distinct quantum-circuit symbol for state-based controlled-gates (including the chameleon gates defined below).
\end{itemize}

An important and interesting feature of this realization is that there is no need to have prior knowledge of the knob state $\rho$ in order to apply a $\mathpzc{U}_{\rho}$ gate. Moreover, the action of the state-based gate is determined according to the given state $\rho$, which itself can be, e.g., the output of some previous step of the algorithm.\\

%%%%%%%%%%%%%%%%%%%%%%%%%%%%%%%%%%%%%%%%%%%%%%%%%%%%%%%%%%%%%%%%
\textit{Chameleon gate.}---The complete effective action of the realization process described in the previous section, without ignoring $O(\delta^{2})$ terms, is given by the following operation:
\begin{align}
\mathcal{C}_{\rho}: 
\begin{cases}
|0 \rangle\langle 0| \otimes \sigma &\to~~ |0 \rangle\langle 0| \otimes \sigma\\
|0 \rangle\langle 1| \otimes \sigma &\to~~ |0 \rangle\langle 1| \otimes \sigma\rho \\
|1 \rangle\langle 0| \otimes \sigma &\to~~ |1 \rangle\langle 0| \otimes \rho\sigma \\
|1 \rangle\langle 1| \otimes \sigma &\to~~ |1 \rangle\langle 1| \otimes \rho \\
\end{cases}
\label{chameleon-gate}
\end{align}
This defines a new type of state-based quantum controlled-operation which we refer to as a ``chameleon gate'' or ``chameleon element'' $\mathcal{C}_{\rho}$. It can be verified easily that
\begin{equation}
\mathcal{C}_{\rho}[\psi_{\delta} \otimes \sigma]= \mathpzc{U}_{\rho} (\psi_{\delta} \otimes \sigma) \mathpzc{U}^{\dag}_{\rho} + O(\delta^{2}).
\end{equation}
In the case the knob state is similar to the quantum state of either the inputs control or target systems, rather than an independent state, the chameleon gate becomes a nonlinear state-based operation on its inputs.
%In the case the action of a chameleon gate is dictated by the very quantum state of either its control or target systems, rather than a fixed given quantum state, it becomes a nonlinear state-based operation on its inputs.

Further, the proposed realization procedure for chameleon gates/elements allows us to lay out a \textit{chameleon-inside} quantum processing unit (QPU) using existing elements for standard gate-based quantum computation, which in principle can be demonstrated in all current approaches for quantum computation. In particular, one can think of in-chip chameleon elements and using them as building blocks of SBQC. 

In the following we demonstrate an application of the chameleon gates for designing SBQC algorithms. We illustrates this through examples. 
\\
%%%%%%%%%%%%%%%%%%%%%%%%%%%%%%%%%%%%%%%%%%%%%%%%%%%%%%%%%%%%%%%%
%\section{State-based quantum computation}
%\label{sec:SBQC}

\textit{State-based quantum computation.---}A (universal) state-based quantum computer uses various quantum states to form suitable chameleon gates for simulation of any quantum evolutions \cite{SBQS}. In this method the given Hamiltonian (which can be even non-Hermitian  \cite{IT-SBQS}) which generates the dynamics is decomposed in terms of a set of quantum \textit{states}---which we refer to as \textit{resource states}. This set of states can be either a set of given known quantum states or can even include the unknown quantum states of the computer (or a section of it) at different past times. 

The simulation is performed by applying chameleon gates, whose actions are determined by quantum states taking part in forming the Hamiltonian. The action of the chameleon gates are controlled by the state of some qubit ancilla. The final result is obtained by post-selection of the result of the measurements on the ancilla. 

In the following we show a step-by-step simulation of the dynamics generated by two classes of Hamiltonians. We first consider a \textit{given} Hamiltonian $H(t)$ which is \textit{known} during the evolution. Next we consider a more general case, where $H$ is a function of the instantaneous or the past states of the target system (which is here the simulator system) and hence it is \textit{a priori} \textit{unknown}. We denote the states of the simulator with ``$\sigma$'' to distinguish it from the states ``$\rho$'' of the resource systems used for SBQC.\\

%%%%%%%%%%%%%%%%%%%%%%%%%%%%%%%%%%%%%%%%%%%%%%%%%%%%%%%%%%%%%%%%
\begin{figure}[tp]
\includegraphics[width=0.78\linewidth]{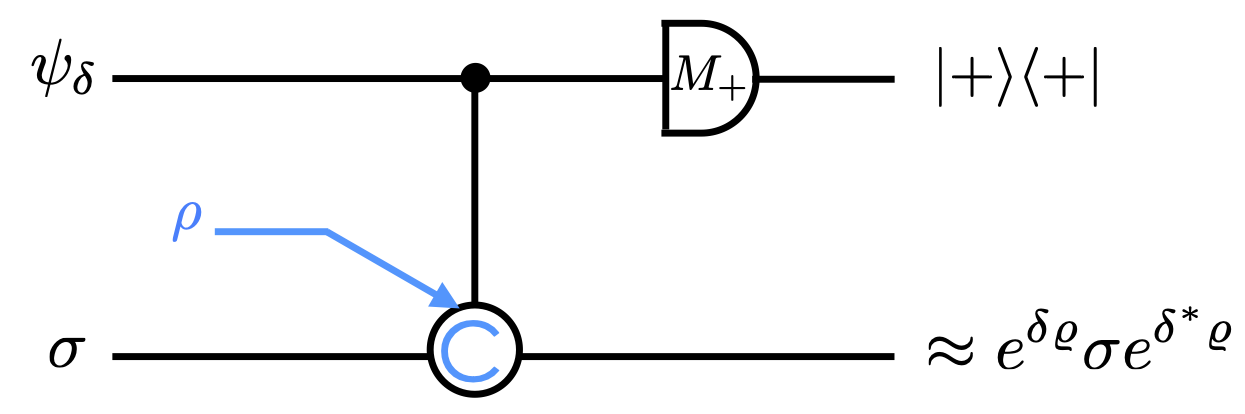}
\caption{SBQC circuit for simulating the dynamics generated by $H=h \rho$ on the simulator state, i.e., $e^{\delta \rho} \sigma e^{\delta^{*} \rho}$.}
\label{fig:SBQS-circuit-H-1}
\end{figure}
%%%%%%%%%%%%%%%%%%%%%%%%%%%%%%%%%%%%%%%%%%%%%%%%%%%%%%%%%%%%%%%%
 
%%%%%%%%%%%%%%%%%%%%%%%%%%%%%%%%%%%%%%%%%%%%%%%%%%%%%%%%%%%%%%%%
 \textit{State-independent Hamiltonians.}---The SBQC method for a \textit{known} Hamiltonian simulation consists of the following steps:\\

\begin{itemize}

\item[(\textbf{I})] \textit{Decomposition of $H$ in terms of quantum states}: A set of density matrices $\{\rho_{j}\}$ is chosen such that we can expand 
\begin{align}
H=\textstyle{\sum}_{j} h_{j} \rho_{j},
\label{H-StateDecomposition}
\end{align}
where $h_{j}$ are scalar coefficients and in general can be complex numbers. The set $\{\rho_{j}\}$ need not be complete or independent and can be chosen based on the ability to prepare particular quantum states in the lab.\\

\item[(\textbf{II})] \textit{Trotter-Suzuki expansion}: The evolution operator, generated by $H$, is given by $U= e^{-i t \sum_{j} h_{j} \rho_{j}}$. Using the Trotter-Suzuki expansion \cite{Lloyd:UQS} we break down $U$ into a concatenation of factors with a single density matrix (i.e., state) $\rho_{j}$ such that
\begin{align}
U = \big(\textstyle{\prod}_{j} e^{\delta_j \rho_{j}}\big)^{n}+O(t^{2}/n),
\label{U-Trotter}
\end{align}
where $\delta_j=-i t h_j/n$, with $n\gg 1$ and $|\delta_j| \ll 1$. This implies that to simulate $U$ it suffices to generate terms of the form $e^{\delta \rho}$.\\

\item[(\textbf{III})] \textit{Exponentiation of density matrix}: To simulate the evolution $e^{\delta \rho}$ on a quantum system with the initial state $\sigma$, we apply the following steps: 

\begin{itemize}

\item[(a)] Prepare a control qubit $\psi_{\delta}$ [Eq. \eqref{control-qubit}] with $\delta=i t h /n$. 

\item[(b)] Apply a chameleon gate $\mathcal{C}_{\rho}$ on the control qubit and the simulator (target) state, leading to 
\begin{align}
\mathcal{C}_{\rho}[\psi_{\delta}\otimes \sigma]=&\, |0 \rangle\langle 0| \otimes  \sigma +  \delta |1 \rangle\langle 0| \otimes  \rho \sigma + \delta^{*} |0 \rangle\langle 1| \otimes \sigma \rho \nonumber\\
&\, + |\delta|^{2} |1 \rangle\langle 1| \otimes  \rho,
\end{align}

\item[(c)] Perform a measurement on the control qubit in the $\{|\pm\rangle\}$ basis (eigenbasis of the $X$-Pauli operator) and post select $|+\rangle$. That is, apply $M_{+}[\circ]= P_{+} \circ P_{+}$, where $P_{+}= |+\rangle\langle +|$. Thus, we obtain
 \begin{align}
M_{+}\big[\mathcal{C}_{\rho}[\psi_{\delta}\otimes \sigma(0)]\big]=&\, |+ \rangle\langle +| \otimes \big(\sigma(0) +  \delta  \rho \sigma+ \delta^{*} \sigma(0) \rho + |\delta|^{2} \rho\big), \nonumber\\
\approx &\, e^{\delta \rho} \sigma(0)\, e^{\delta^{*} \rho}.
\end{align}
Figure \ref{fig:SBQS-circuit-H-1} shows the corresponding SBQC circuit.

\end{itemize}

\end{itemize}

It is evident that to go further in time, we need to repeat similar simulation steps applied on the output simulator state of this step.\\

%%%%%%%%%%%%%%%%%%%%%%%%%%%%%%%%%%%%%%%%%%%%%%%%%%%%%%%%%%%%%%%%
\begin{figure}[tp]
\includegraphics[width=\linewidth]{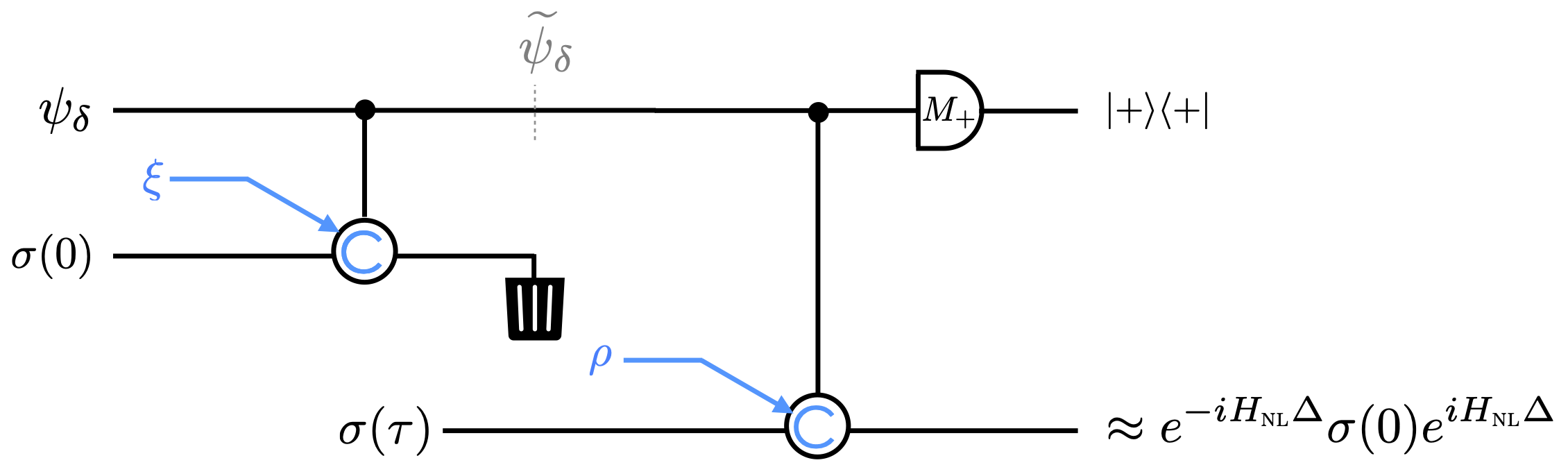}
\caption{State-based quantum circuit for simulation of the dynamics generated by the nonlinear Hamiltonian \eqref{NLH}. For the relation between the parameters $\delta$ and $\Delta$ see the main text.}
\label{fig:nonlin-H-SBQC}
\end{figure}
%%%%%%%%%%%%%%%%%%%%%%%%%%%%%%%%%%%%%%%%%%%%%%%%%%%%%%%%%%%%%%%%

%%%%%%%%%%%%%%%%%%%%%%%%%%%%%%%%%%%%%%%%%%%%%%%%%%%%%%%%%%%%%%%%
\textit{State-dependent Hamiltonians}.---Consider a class of state-dependent Hamiltonians $H(t)$ similar to the one in Eq. \eqref{H-StateDecomposition}
with the following specific form for state-dependent couplings:
\begin{align}
h_{j}\big(\sigma(t-\tau)\big) = c_{j}\mathrm{Tr}[\xi_{j}\sigma(t-\tau)],
\label{hNL}
\end{align}
where $c_{j}$ is a constant real number, $\xi_{j}$ is a given constant trace-unity positive operator, and $\sigma(t-\tau)$ is the delayed state of the simulator system at $t-\tau$. To simulate the dynamics generated by such Hamiltonians using the Trotter-Suzuki expansion, it suffices to simulate the dynamics generated by terms of the form
\begin{align}
H_{\textsc{nl}}=c\,\mathrm{Tr}[\xi \sigma(t-\tau)] \,\rho.
\label{NLH}
\end{align}

For simulating such delay systems, we need to know not only the initial state $\sigma(0)$ but also $\sigma(t)$  for all times $s$ in the interval $0\leqslant s \leqslant \tau$. With this information, we perform the following steps to simulate the evolution of the system which is in the state $\sigma(\tau)$ into the next time step $\tau+\Delta$, where $\Delta \ll \tau$:

\begin{itemize}

\item[(\textbf{a})] Simulation steps are akin to the case of state-independent Hamiltonian, but with a subtle difference in its (\textbf{III}) (a) step. Here the coefficients of the needed state of the control qubit depend on the state of the simulator at past times: $|\tilde{\psi}_{\delta}\rangle=|0 \rangle + \delta \,\mathrm{Tr}[\xi \sigma(0)] |1 \rangle$, where $\delta= -i c \Delta$ and $|\delta| \ll 1$. To prepare such a control qubit state, we perform the following steps:
 
\begin{itemize}

\item[(i)] Prepare a control qubit in the state $\psi_{\delta}$.

\item[(ii)] Apply a chameleon gate $\mathcal{C}_{\xi}$ on the control qubit and a copy of the simulator state at the initial time:
\begin{align}
\mathcal{C}_{\xi}[\psi_{\delta} \otimes  \sigma(0)] = & \, |0 \rangle\langle 0| \otimes  \sigma(0) +  \delta |1 \rangle\langle 0| \otimes  \xi \,\sigma(0) \\
&\, + \delta^{*} |0 \rangle\langle 1| \otimes \sigma(0)\, \xi + |\delta|^{2} |1 \rangle\langle 1| \otimes  \xi. \nonumber
\label{post-selection}
\end{align}

\item[(iii)] Trace out the target system to find an updated qubit state
\begin{align}
\mathrm{Tr}_{\textsc{t}}\left[\mathcal{C}_{\xi}[\psi_{\delta} \otimes  \sigma(0)] \right] = \tilde{\psi}_{\delta} + O(\delta^{2}),
\end{align}
where $\tilde{\psi}_{\delta}= |\tilde{\psi}_{\delta}\rangle\langle \tilde{\psi}_{\delta}|$.

\end{itemize}

Now that we have prepared the suitable control qubit state, we follow similar steps as the state-independent case for the rest of the simulation process.\\

\item[(\textbf{b})] Apply a chameleon gate $\mathcal{C}_{\rho}$ on the updated control qubit and a copy of the simulator state at time $\tau$, i.e., $\sigma(\tau)$:
\begin{align}
\mathcal{C}_{\rho}&\big[\tilde{\psi}_{\delta} \otimes \sigma(\tau)\big] = |0 \rangle\langle 0|\otimes \sigma(\tau) \nonumber\\
&+\big(\delta\mathrm{Tr}[\xi \sigma_0] |1 \rangle\langle 0|\otimes \rho \sigma(\tau)+ \mathrm{h.c.}\big)+ O(\delta^{2}).
\end{align}
\item[(\textbf{c})] Perform a measurement on the control qubit in the $\{|\pm\rangle\}$ basis and post select $|+\rangle$:
\begin{align}
M_+\big[\mathcal{C}_{\rho}&[\tilde{\psi}_{\delta} \otimes \sigma(\tau)]\big]\nonumber\\
&=|+\rangle\langle +|\otimes \sigma(\tau)+ \delta\, \mathrm{Tr}[\xi \sigma_0][\rho,\sigma(\tau)] + O(\delta^{2}) \nonumber\\
& \approx  |+\rangle\langle +|\otimes e^{-i H_{\textsc{nl}} \Delta} \sigma(\tau) e^{i H_{\textsc{nl}}\Delta}.
\end{align}
See Fig. \ref{fig:nonlin-H-SBQC} for the related SBQC circuit. 
\end{itemize}

\textit{Competing Interest.---}The author are inventors on a pending patent application related to the methods described in this manuscript.

%%%%%%%%%%%%%%%%%%%%%%%%%%%%%%%%%%%%%%%%%%%%%%%%%%%%%%%%%%%%%%%%

%%%%%%%%%%%%%%%%%%%%%%%%%%%%%%%%%%%%%%%%%%%%%%%%%%%%%%%%%%%%%%%%

\end{document}